\documentclass[aps,preprintnumbers,nofootinbib,superscriptaddress, 11pt]{revtex4}
\usepackage[hypertex]{hyperref}
\usepackage[dvipdfmx]{graphicx} 
\usepackage{amsmath,amssymb,amsfonts,bm,cancel} 

\def\a{\alpha} \def\b{\beta}   \def\d{\delta} \def\D{\Delta}    \def\th{\theta}    \def\L{\Lambda} \def\m{\mu} \def\n{\nu}     \def\r{\rho} \def\s{\sigma}  \def\t{\tau}       

\def\dg{\dagger}  \def\nn{\nonumber}

\newcommand{\meV}{ \, {\rm meV} }   \newcommand{\MeV}{\, {\rm MeV} } \newcommand{\GeV}{\, {\rm GeV} } \newcommand{\TeV}{\, {\rm TeV} }
\newcommand{\PeV}{\, {\rm PeV} }

\newcommand{\lsp}{ \left ( } \newcommand{\rsp}{ \right ) }

\def\abs#1{\left| #1\right|}
 
  \newcommand{\Det}{{\, \rm Det  \,}}

\newcommand{\row}[2]{ \begin{pmatrix}  #1 & #2   \end{pmatrix}  }
\newcommand{\column}[2]{ \begin{pmatrix}  #1 \\ #2 \\  \end{pmatrix} }

\newcommand{\diag}[2]{ \begin{pmatrix}  #1 & 0 \\ 0 & #2 \\   \end{pmatrix}  }

\newcommand{\Diag}[3]{ \begin{pmatrix} #1 & 0 & 0 \\ 0 & #2 & 0 \\ 0 & 0 & #3 \\\end{pmatrix}}

\begin{document}


\title{\large Allowed region for the (second) lightest mass $M_{1,2}$ of 
right-handed neutrino $\nu_{R 1,2}$ with $SO (10)$-inspired relations and sequential dominance }
\preprint{STUPP-23-268}
\author{Masaki J. S. Yang}
\email{mjsyang@mail.saitama-u.ac.jp}
\affiliation{Department of Physics, Saitama University, 
Shimo-okubo, Sakura-ku, Saitama, 338-8570, Japan}



\begin{abstract} 

In this paper, we evaluate the (second) lightest mass $M_{1,2}$ of right-handed neutrino $\nu_{R1,2}$  in the type-I seesaw mechanism with $SO(10)$-inspired relations that predicts an almost massless neutrino $m_{1 \, \rm or \, 3} \sim 0$. 
By chiral perturbative treatment, the masses $M_{1,2}$ are expressed as $M_{1} = m_{D1}^{2}/m_{11} , \, M_{2} = m_{D2}^{2} m_{11} / (m_{11} m_{22} - m_{12}^{2})$ with the mass matrix of left-handed neutrinos $m$ in the diagonal basis of the Dirac mass matrix $m_{D}$.

Assuming $m_{Di}$ and the unitary matrix $V$ in the singular value decomposition $(m_{D})_{ij} = \sum_{k} V_{ik} m_{D k} U^{\dagger}_{kj}$ are close to observed fermion masses and the CKM matrix, we obtain expressions of $M_{1,2}$ by parameters in the low energy and unknown phases. 
As a result, for $m_{D1} \simeq 0.5$ MeV and $m_{D2} \simeq 100$ MeV, 
their allowed regions are found to be
$M_{1}^{\rm NH} \simeq 3 \times 10^{4 - 6}$ GeV and  
$M_{2}^{\rm NH} \simeq 3 \times 10^{6 - 8}$ GeV in the NH, 
$M_{1}^{\rm IH} \simeq 5 \times 10^{3 - 4}$ GeV and 
$M_{2}^{\rm IH} \simeq 4 \times 10^{8 - 9}$ GeV in the IH. 
These upper and lower bounds are proportional to $m_{Di}^{2}$ or $m_{D1} m_{D2}$.

\end{abstract} 

\maketitle

\section{Introduction}

The type-I seesaw mechanism \cite{Minkowski:1977sc,GellMann:1980v,Yanagida:1979as, Mohapatra:1979ia} remains one of the most important subjects in particle physics.
For the three mass matrices in this mechanism, texture studies 
have been extensively investigated  
\cite{Fritzsch:1999ee, Xing:2015fdg, Xing:2020ijf, Xing:2020ald, Xing:2022uax}. 
For example, the sequential dominance (SD) \cite{King:1998jw, King:1999cm, King:1999mb, King:2002nf, Antusch:2004gf, Antusch:2004re,  Antusch:2006cw, Antusch:2010tf, King:2005bj, Antusch:2007dj, Antusch:2011ic, Antusch:2013wn, Bjorkeroth:2014vha}
attempt to explain the large mixings of the neutrino oscillation 
by a very large mass $M_{3}$ of the heaviest right-handed neutrino $\n_{R3}$.
In the limit of $M_{3} \to \infty$, a neutrino approaches to be massless. 
It has also been argued that such a light neutrino is associated with an approximate lepton number symmetry \cite{Wyler:1982dd, Petcov:1984nz,  Branco:1988ex, Kersten:2007vk, Adhikari:2010yt}. 
These lepton number symmetries $U(1)_{L}^{3} \times U(1)_{R}^{3}$ can be reinterpreted as kind of chiral symmetries.

Meanwhile, chiral perturbative treatment by singular values $m_{Di}$ of the Dirac mass matrix $m_{D}$ has made it possible to calculate three right-handed neutrino masses $M_{i}$ 
\cite{Branco:2002kt, Akhmedov:2003dg, Davidson:2003cq}. 
These results are used in the evaluation of $N_{2}$ leptogenesis \cite{DiBari:2005st, Vives:2005ra} based on $SO(10)$-inspired mass relations \cite{DiBari:2008mp, Bertuzzo:2010et,DiBari:2010ux, DiBari:2013qja, DiBari:2014eqa, DiBari:2014eya, DiBari:2017uka, Chianese:2018rnq} (and for a review, \cite{DiBari:2021fhs}). 
In this treatment, input parameters are the mass matrix of light neutrinos $m_{\n}$, the singular values $m_{Di}$ of the Dirac mass matrix $m_{D}$, and the left-handed unitary matrix $V$ in the singular value decomposition $m_{D} = V m_{D}^{\rm diag} U^{\dg}$.
Thus, by assuming $m_{Di}$ and $V$ to be about the observed fermion masses and the CKM matrix $V_{\rm CKM}$, we can understand analytic dependences of $M_{i}$ on the low energy observables in some grand unified theories. 
 
In this paper, as a simplified problem, 
we evaluate the lighter masses $M_{1,2}$ when the lightest neutrino mass $m_{1 \rm \, or \, 3}$ is very small.
In the parameter region where the chiral perturbative expansion is valid, the limit $m_{1 \rm \, or \, 3} \to 0$ corresponds to the sequential dominance with $M_{3} \to \infty$. 
We will analyze the allowed regions for the lighter masses $M_{1,2}$ in such a model. 
This result would be valid for discussions of cosmology, such as leptogenesis, because it is encompassed by the results of some grand unified theories with seesaw mechanisms.

This paper is organized as follows. The next section gives a review of the chiral perturbative diagonalization of $M_{R}$ and the sequential dominance  in the type-I seesaw mechanism. 
In Sec.~3, we evaluate $M_{1,2}$ with $SO(10)$-inspired relations and  the sequential dominance. 
The final section is devoted to a summary.

\section{Chiral perturbative diagonalization of $M_{R}$}

In this section, we review the chiral perturbative diagonalization of the mass matrix of right-handed neutrinos $M_{R}$ \cite{Akhmedov:2003dg, DiBari:2014eya, DiBari:2017uka, Yang:2022bex, Yang:2023ixv}. 
The Dirac mass matrix $m_{D}$ and the mass matrix of neutrinos $m_{\n} = U_{\rm MNS} m_{\n}^{\rm diag} U_{\rm MNS}^{T}$ are defined in a basis where the Yukawa matrix of charged leptons $Y_{e}$ and $M_{R}$ are diagonal. 
Here, $U_{\rm MNS}$ is the MNS matrix.
By the singular value decomposition $m_{D} = V m_{D}^{\rm diag} U^{\dg}$ with unitary matrices $V$ and $U$, 
the type-I seesaw mechanism in the diagonal basis of $m_{D}$ is represented as
\begin{align}
  m_{D}^{\rm diag} V^{T} m_{\n}^{-1} V m_{D}^{\rm diag } 
& = U^{T} M_{R}^{\rm diag } U  \, .
\end{align}
%
By defining the mass matrix $V^{T} m_{\n}^{-1} V \equiv m^{-1}$ (and $V^{\dg} m_{\n} V^{*} = m $) in this basis,  matrix elements of $M_{R}$ are simply written by $(M_{R})_{ij} \equiv (U^{T} M_{R}^{\rm diag } U)_{ij} = m_{D i } (m^{-1})_{ij} m_{Dj}$.  
Here, singular values $m_{Di}$ of $m_{D}$ are assumed to be sufficiently hierarchical ($m_{D3} \gg m_{D2} \gg m_{D1}$).

For an invertible $m$ (i.e., the lightest mass $m_{1 \, \rm or \, 3 }$ is finite), 
$M_{R}$ is diagonalized chiral perturbatively.
Once the heaviest right-handed neutrino $\n_{R3}$ is integrated out, 
the mass matrix $M_{R0}$ of the remaining two lighter generations is
\begin{align}
M_{R0} 
 & = 
\begin{pmatrix}
m_{D1}^{2} (m^{-1})_{11} & m_{D1} m_{D2} (m^{-1})_{12} \\
 m_{D1} m_{D2}  (m^{-1})_{12}  & m_{D2}^{2} (m^{-1})_{22} 
\end{pmatrix} \\
& - {1\over m_{D3}^{2} (m^{-1})_{33}} m_{D3}^{2}
\column{m_{D1}  (m^{-1})_{13} }{m_{D2} (m^{-1})_{23} } \otimes
\row{m_{D1}  (m^{-1})_{13} }{m_{D2}  (m^{-1})_{23} } \\
 & = {1\over m_{11} m_{22} - m_{12}^{2}}
\diag{m_{D1}}{m_{D2}}
\begin{pmatrix}
m_{22} & - m_{12} \\ - m_{12} & m_{11}
\end{pmatrix} 
\diag{m_{D1}}{m_{D2}} . 
\label{MR0}
\end{align}
%
This means that to integrate out $\n_{R3}$ simply yields
the seesaw mechanism of the remaining two generations.

By integrating out the second heaviest neutrino $\n_{R2}$ again, 
$m^{-1}$ becomes the  ``inverse of $1\times1$ matrix'' $1 / m_{11}$, 
and the diagonalized elements $M_{i}$ of the  mass matrix $M_{R}$ are found to be
\begin{align}
(M_{1} \, , \, M_{2} \, , \, M_{3}) \simeq 
 ( {m_{D1}^{2} \over m_{11} } \, ,  \,   {m_{D2}^{2} \, m_{11} \over m_{11} m_{22} - m_{12}^{2} } \, , \,  m_{D3}^{2} (m^{-1})_{33} ) \, . 
\label{Mi}
\end{align}
%
Strictly the singular values of $M_{R}$ are the absolute values of Eq.~(\ref{Mi}).
This is a kind of chiral perturbative expansion because $M_{i}$ becomes zero in the limit of $m_{Di} \to 0$.

The diagonalization matrix of $M_{R}$ associated with the integration out of heavy states is 
\footnote{ The 1-3 element of $U$ is not a misprint, but due to the sum of two terms. 
 This makes $U$ not look like a unitary matrix, because $m_{D1}/m_{D3}$ behaves as a second-order perturbation.}
\begin{align}
U &\simeq 
\begin{pmatrix}
 1 &   {m_{D1} \over m_{D2}} {m_{12}^{*} \over m_{11}^{*}} & 0 \\
- {m_{D1} \over m_{D2}}   {m_{12} \over m_{11} } & 1 & 0 \\
 0 & 0 & 1 \\
\end{pmatrix} 
\begin{pmatrix}
 1 & 0 & - {m_{D1} \over m_{D3}} {(m^{-1})^{*}_{13} \over (m^{-1})^{*}_{33} } \\
 0 & 1 & - {m_{D2} \over m_{D3}}  { (m^{-1})^{*}_{23} \over (m^{-1})^{*}_{33} } \\
  {m_{D1} \over m_{D3}}  { (m^{-1})_{13} \over (m^{-1})_{33} } & {m_{D2} \over m_{D3}} { (m^{-1})_{23} \over (m^{-1})_{33} } & 1 \\
\end{pmatrix}  \\ 
& \simeq 
\begin{pmatrix}
 1 &   {m_{D1} \over m_{D2}} {m_{12}^{*} \over m_{11}^{*}} &  {m_{D1} \over m_{D3}} {m_{13}^{*} \over m_{11}^{*}}  \\
- {m_{D1} \over m_{D2}}   {m_{12} \over m_{11} } & 1 &  - {m_{D2} \over m_{D3}}  { (m^{-1})^{*}_{23} \over (m^{-1})^{*}_{33} }  \\
  {m_{D1} \over m_{D3}}  { (m^{-1})_{13} \over (m^{-1})_{33} } & {m_{D2} \over m_{D3}} { (m^{-1})_{23} \over (m^{-1})_{33} } & 1 \\
\end{pmatrix} \, . 
\end{align}
This procedure perturbatively ``solves'' all six constraints of the type-I seesaw mechanism \cite{DiBari:2014eya, DiBari:2017uka}.
In other words, $U$ and $M_{i}$ are determined by input parameters $m_{\n}, m_{Di}$ and $V$ \cite{Davidson:2001zk}.

For the chiral perturbative diagonalization to be a good description, 
$U$ must not have a large mixing and the following conditions are required;  
\begin{align}
{m_{D1} \over m_{D2}}  \abs{m_{12} \over m_{11} } \, , ~   {m_{D1} \over m_{D3}} \abs { (m^{-1})_{13} \over (m^{-1})_{33} } \, , ~  {m_{D2} \over m_{D3}} \abs{ (m^{-1})_{23} \over (m^{-1})_{33} } \lesssim 0.1 \, . 
\end{align}
Since elements $m_{ij}$ are expected to be comparable unless $m_{\n}$ and $V$ have special forms, this perturbative treatment is valid in a wide parameter region.

\subsection{Sequential dominance by the limit of $m_{1 \, \rm or \, 3} \to 0$}

The above discussion requires that $m_{\n}$ have an inverse matrix.  
Let us examine the behavior of the limit $m_{\rm 1\, or \, 3} \to 0$ 
(This is corresponds to performing a chiral perturbative expansion by $m_{\rm 1\, or \, 3}$). 
There are two limits in which the matrix $m_{\n} = m_{D} M_{R}^{-1} m_{D}^{T}$ becomes singular; 
\begin{enumerate}
\item The rank of $m_{D}$ approaches two.
\item $M_{3} \to \infty$.
\end{enumerate}

In the case of 1, (almost) zero singular values of $m_{D}$ and $m_{\n}$ yields
a chiral symmetry associated with the lepton number \cite{Branco:1988ex, Kersten:2007vk, Adhikari:2010yt, Yang:2022yqw}; 
\begin{align}
R \, m_{D} = m_{D} \, ,  ~~~
R \, m_{\n} = m_{\n} \, R = m_{\n} ,  ~~~
R \equiv {\rm diag} (e ^{ i  \a} \, , 1 \, , 1 ) . 
 \label{chiralR}
\end{align}
The following shows that $R$ for $m_{\n}$ and $m_{D}$ is identical.
For each singular value decomposition 
$m_{\n} \equiv U_{\rm MNS} m_{\n}^{\rm diag} U_{\rm MNS}^{T}$ and 
$m_{D} = V m_{D}^{\rm diag} U^{\dg}$, 
the seesaw mechanism is written by 
\begin{align}
\Diag{0}{m_{2}}{m_{3}}
= 
U_{\rm MNS}^{\dg} V \Diag{0}{m_{D2}}{m_{D3}} U^{\dg} (M_{R}^{\rm diag})^{-1} U^{*} \Diag{0}{m_{D2}}{m_{D3}}  V^{T} U_{\rm MNS}^{*}\, .
\end{align}
The first column of $U_{\rm MNS}$ and $V$ must match in order for the singular value decomposition (by $U_{\rm MNS}^{\dg} V$) to close in the 2-3 subspace. 
Thus, the eigenvectors of $U_{\rm MNS}$ and $V$ in the massless direction must coincide, 
and Eq.~(\ref{chiralR}) holds simultaneously in a proper basis. 

For the normal hierarchy (NH) $m_{1} = 0$ and the inverted hierarchy (IH) $m_{3} = 0$, 
the form of $m = V^{\dg} m_{\n} V^{*}$ in this basis is respectively
\begin{align}
m = 
\begin{pmatrix}
0 & 0 & 0 \\
0 & * & * \\
0 & * & * \\
\end{pmatrix}
~~ {\rm or} ~~
\begin{pmatrix}
 * & * & 0 \\
 * & *  & 0 \\
0 & 0 & 0 \\
\end{pmatrix} ,
\label{8}
\end{align}
where $*$ denotes appropriate matrix elements.
That is, $m_{1i}$ or $m_{3i} =0$ is a consequence of rank $m_{D} = 2$ (for $\det M_{R} \neq 0$). 
In this case the perturbative diagonalization for $M_{R}$ breaks down \cite{Yang:2023ixv}, 
because the cancellation of the approximate chiral symmetry between $m_{D}$ and $m_{\n}^{-1}$ leads to a matrix $M_{R}$ without the chiral symmetry. 
Chiral perturbative treatment can be possible in the remaining two generations. 
However, in this situation, the unitary matrix $V \sim U_{\rm MNS}$ of the Dirac mass matrix is far from the CKM matrix. Specifically,
\begin{align}
V^{\rm NH} \sim 
\begin{pmatrix}
- {2 / \sqrt 6} & * & * \\
{1 / \sqrt 6} & * & * \\
{1 / \sqrt 6} & * & * \\
\end{pmatrix} , ~~~
V^{\rm IH} \sim 
\begin{pmatrix}
 * & * & 0 \\
 * & * & - 1/ \sqrt 2 \\
 * & * & 1/ \sqrt 2 \\
\end{pmatrix} ,
\end{align}
for the case of NH and IH. 
Thus, there is no guideline for $m_{Di}$ and $V$ from unified theories. 

On the other hand, the second case $M_{3} \to \infty$ can be treated by chiral perturbative expansion and it corresponds to the sequential dominance \cite{King:1998jw, King:1999cm, King:1999mb}. 
Since several parameters become unphysical as the heaviest neutrino $\n_{R3}$ is decoupled, the seesaw mechanism is simplified and analysis becomes easier in this limit.

\subsection{Behavior of $M_{R}$ in the limit of $m_{\rm 1 \, or \, 3} \to 0$}

Here, we explore the chiral perturbative behavior of $M_{R}$ 
in the limit of $m_{\rm 1 \, or \, 3} \to 0$. 
A misalignment of diagonalization $W$ between $m_{\n}$ and $m_{D}$ is defined as $V^{\dg} U_{\rm MNS} \equiv W \equiv (\bm w_{1}, \bm w_{2}, \bm w_{3})$ by 3-dimensional vectors $\bm w_{i}$. 
The spectral decomposition 
$m^{-1} = V^{T} m_{\n}^{-1} V = W^{*} m_{\n}^{\rm diag -1} W^{\dg}$ is written by
\begin{align}
(m^{-1})^{*} = {1\over m_{1}} \bm w_{1} \otimes \bm w_{1}^{T}
+  {1\over m_{2}} \bm w_{2} \otimes \bm w_{2}^{T}
+  {1\over m_{3}} \bm w_{3} \otimes \bm w_{3}^{T} \, . 
\label{minvstar}
\end{align}
In Eq.~(\ref{minvstar}), the rank one matrix diverging in the limit of $m_{1 \, \rm or \, 3} \to 0$ contributes to $M_{3}$.
%
%
In particular,  
\begin{align}
(m^{-1})_{33}^{*} &= {W_{31}^{2}\over m_{1}} +  {W_{32}^{2} \over m_{2}} 
+  {W_{33}^{2}\over m_{3}} 
\simeq  {W_{31}^{2}\over m_{1}} \, {\rm or} \,  {W_{33}^{2}\over m_{3}} \, . 
\label{minv33}
\end{align}
If the matrix $V$ is similar to $V_{\rm CKM}$, 
$W$ has large mixings and 
matrix elements of $W$ are comparable. 
Then this approximation holds with $m_{1 \, \rm or \, 3} / m_{2} \lesssim 0.1$, and
it is valid when $m_{1} \lesssim 1 \, \meV$ in NH and $m_{3} \lesssim 5 \, \meV$ in IH. 
In this case the heaviest singular value $M_{3}$ is evaluated as 
\begin{align}
M_{3} \simeq m_{D3}^{2} (m^{-1})_{33}
\simeq {m_{D3}^{2}  \over m_{1}}  \abs{W_{31}^{*}}^{2}
~~ {\rm or} ~~ {m_{D3}^{2} \over m_{3}} \abs{W_{33}^{*}}^{2}  \, .  
\label{13}
\end{align}
This situation will be realized by a sufficiently heavy $M_{3}$ in some GUT models with strong hierarchy. 

These chiral relations of $M_{i}$ are expected to hold at a higher energy scale because the Yukawa interactions of the first and second generations hardy receive the renormalization. 
Only when $m_{D3}$ is about as large as the top quark mass $m_{t}$, it receives a quantum correction of about 10 \% \cite{Xing:2007fb}.
Furthermore, the renormalization of gauge interactions in the Standard Model cancels out in the numerator and denominator, the expressions for $M_{1,2}$ are rather insensitive to quantum corrections. 

\section{The (second) lightest  mass $M_{1,2}$ of right-handed neutrino $\n_{R 1,2}$ with $SO(10)$-inspired relations and sequential dominance}

Here we evaluate the lighter masses $M_{1,2}$ and its allowed regions  with $SO(10)$-inspired relations and sequential dominance.
First of all, the singular value decompositions of the Yukawa matrices of leptons in a general basis are defined as 
\begin{align}
Y_{\n}^{0} \equiv V_{\n} Y_{\n}^{\rm diag} U_{\n}^{\dg} \, , ~~~
Y_{e}^{0} \equiv V_{e} Y_{e}^{\rm diag} U_{e}^{\dg} \, . 
\end{align}
On the other hand, the mass matrices $m_{D}$ and $m_{\n}$ are defined in the diagonal basis of $Y_{e}$; 
\begin{align}
Y_{e} = Y_{e}^{\rm diag} \, , ~~
m_{D} = V m_{D}^{\rm diag} U^{\dg} \, , ~~
m_{\n} = U_{\rm MNS} m_{\n}^{\rm diag} U_{\rm MNS}^{T} \, . 
\end{align}
From this, $V = V_{e}^{\dg} V_{\n}$ and $U_{\rm MNS}$ implicitly involve $V_{e}$.

With some grand unified theories in mind, the following $SO(10)$-inspired relations are assumed.
\begin{itemize}
\item[1.]  $m_{D1} \simeq m_{u}(\L_{\rm GUT}) \simeq  m_{e}(\L_{\rm GUT}) \simeq 0.5 \MeV$,
$m_{D2} \simeq m_{c}(\L_{\rm GUT}) \sim  m_{\m}(\L_{\rm GUT}) \sim 100 \MeV$ \cite{Xing:2007fb}. 
\item[2.] 
The unitary matrix $V$ of $m_{D}$ is dominated by 1-2 mixing, and its mixing angle is at most about the Cabibbo angle $\th_{C}$.
\end{itemize}
Specifically, $V$ is expressed by, 
\begin{align}
V \simeq 
\Diag{e^{i \s_{1}}}{e^{i \s_{2}}}{e^{i \s_{3}}}
\begin{pmatrix}
c_{\th} & s_{\th} e^{i \phi}  &   O(10^{-3}) \\
- s _{\th} e^{-i\phi} & c_{\th} &  O(10^{-2}) \\
 O(10^{-3}) &  O(10^{-2}) & 1 
\end{pmatrix} , 
\end{align}
where $s_{\th} \equiv \sin \th \lesssim 0.2 , c_{\th} \equiv \cos \th \sim 1$ and $\phi, \s_{i}$ are unknown phases. 
The signs of $s_{\th}, c_{\th}$ are absorbed into the phases, and 
$\th$ is restricted to the first quadrant $0 \leqq \th \leqq \pi/2$.
Unknown phases $\r_{i}$ from $V_{\n}$ are omitted because 
their contributions $m_{ij} \to e^{i \r_{i}} m_{ij} e^{i \r_{j}}$ are unphysical.

Since these assumptions eliminate the possibility of identical chiral symmetry such as Eq.~(\ref{8}), we consider only SD in this situation. 
From $|W_{3i}| \simeq |U_{\t i}|$, the absolute value of $M_{3}$~(\ref{13}) is hardly affected by $V$;  
\begin{align}
M_{3} \simeq m_{D3}^{2} (m^{-1})_{33}
\simeq {m_{D3}^{2}  \over m_{1}}  \abs{U_{\t 1}}^{2}
~~ {\rm or} ~~ {m_{D3}^{2} \over m_{3}} \abs{U_{\t3}}^{2}  \, . 
\end{align}

To express $M_{1,2}$ by low energy parameters, 
$m_{11}$ is written by matrix elements of $m_{\n}$ with $m_{\a\b} = \sum_{i} m_{i} U_{\a i} U_{\b i}$ as
\begin{align}
m_{11} &\simeq (V_{11}^{*})^{2} m_{ee} + 2 V_{11}^{*} V_{21}^{*} m_{e \m} + (V_{21}^{*})^{2} m_{\m \m} \, , \\
& =  c_{\th}^{2} e^{- 2 i \s_{1} }  m_{ee} - 2 s_{\th} c_{\th} e^{i (\phi - \s_{1} -  \s_{2})} m_{e\m} + s_{\th}^{2} e^{2 i (\phi -  \s_{2})} m_{\m\m}  \, . 
\end{align}
Since the final result is an absolute value, the extra phase can be removed by multiplying by $e^{2 i \s_{1}}$;
\begin{align}
|m_{11}| 
& \simeq |c_{\th}^{2} m_{ee} - 2 s_{\th} c_{\th} e^{i \b} m_{e\m} + s_{\th}^{2} e^{2 i \b} m_{\m\m} | \, ,
\end{align}
where $\b \equiv \phi + \s_{1} -  \s_{2}$ 
is the only unknown phase that cannot be determined from phenomenology.

Besides, using elements $W_{ij}$ and $m_{i}$, we obtain
\begin{align}
m_{11}^{\rm NH}  \simeq W_{12}^{2} m_{2}  + W_{13}^{2} m_{3}  \, ,   ~~~
m_{11}^{\rm IH} \simeq W_{11}^{2} m_{1} + W_{12}^{2} m_{2}  \, .
\end{align}
By uniform redefinition of phases $e^{ i \s_{1} } W_{1i} = c_{\th} U_{ei} - s_{\th} e^{i \b } U_{\m i}$ and PDG parameterization of $U_{\rm MNS}$, these matrix elements are rewritten as 
\begin{align}
e^{i \s_{1}} W_{11} & \simeq c_{\theta } c_{12} c_{13} + s_{\theta } e^{i \b } ( c_{23} s_{12} + c_{12} s_{13} s_{23} e^{i \d}) \, ,  \\
e^{-i\a/2} e^{i \s_{1}} W_{12} & \simeq  c_{\theta } s_{12} c_{13} - s_{\theta } e^{i \b } (c_{12} c_{23} - s_{12} s_{13} s_{23} e^{i\d}) \, , \\
e^{i \s_{1}} W_{13} & \simeq  c_{\theta } s_{13} e^{-i \delta } - s_{\theta } e^{i \b} c_{13} s_{23} \, ,
\end{align}
where $c_{ij} \equiv \cos \th_{ij}$ and $s_{ij} \equiv \sin \th_{ij}$. 
In the limit of $M_{3} \to \infty$, the only Majorana phase $\a$ is defined as associated with $m_{2}$.

The latest global fit (without Super-Kamiokande) is used for input parameters \cite{Gonzalez-Garcia:2021dve}. 
The mass differences are 
\begin{align}
\D m^{2}_{21} = 74.2 \meV^{2} \, , ~~ \D m^{2}_{31} = 2515 \meV^{2} \, , ~~ 
\D m^{2}_{32} = - 2498 \meV^{2} \, , ~~ 
\end{align}
where $\D m^{2}_{31} > 0$ for NH and $\D m^{2}_{32} < 0$ for IH. 
The three mixing angles are
\begin{align}
& \sin^{2} \th_{12}^{\rm NH} = 0.304 \, , ~~ \sin^{2} \th_{23}^{\rm NH} = 0.573 \, , ~~~ \sin^{2} \th_{13}^{\rm NH} = 0.0222 \, , \\
& \sin^{2} \th_{12}^{\rm NH} = 0.304 \, , ~~ \sin^{2} \th_{23}^{\rm NH} = 0.578 \, , ~~~ \sin^{2} \th_{13}^{\rm NH} = 0.0224 \, . 
\end{align}
In this paper, only best-fit values are used to grasp behaviors of $M_{1,2}$. 
The errors in chiral perturbative expansion of $m_{Di}$ are $O(m_{Di}^{2} / m_{Dj}^{2}) \lesssim $ 1 \% \cite{Yang:2024xgi}, which is smaller than that of neutrino oscillation experiments. 
There are also another errors of about 4 \% due to the neglect of $V_{cb}, V_{ts} \sim 0.04$.
However, the largest source of errors come from
the neglect of $m_{1 \, \rm or \, 3}/ m_{2}$. 

\subsection{Inverted hierarchy}

We start the analysis of IH first because it is easier than NH.
From the triangular inequality 
$|z_{1}| - |z_{2}| \leqq | z_{1} + e^{i \a} z_{2}| \leqq |z_{1}| + |z_{2}|$
for complex numbers $z_{1,2}$,
\begin{align}
m_{1} |W_{11}|^{2} - m_{2} |W_{12}|^{2} \leqq  |m_{11}^{\rm IH}| \leqq m_{1} |W_{11}|^{2} + m_{2} |W_{12}|^{2} \, . 
\end{align}
Since $W_{11}$ and $W_{12}$ are functions of $\b$ and $\d$, 
inequality is not used anymore.

The upper bound is evaluated as 
\begin{align}
& m_{1} |W_{11}|^{2} + m_{2} |W_{12}|^{2}  \\
& \simeq m_{1} \abs{c_{\theta } c_{12} c_{13} + s_{\theta } e^{i \b } ( c_{23} s_{12} + c_{12} s_{13} s_{23} e^{i \d})}^{2} 
+ m_{2} \abs{ c_{\theta } s_{12} c_{13} - s_{\theta } e^{i \b } (c_{12} c_{23} - s_{12} s_{13} s_{23} e^{i\d})}^{2} \nn \\ 
& = 
c_{13}^2 c_{\th}^2  ( m_1c_{12}^2 + m_2 s_{12}^2 ) 
+2 c_{13} c_{\th} s_{\th} [s_{13} s_{23} ( m_1 c_{12}^2 + m_2 s_{12}^2 ) \cos (\beta + \delta ) + c_{12} c_{23} (m_1-m_2 ) s_{12} \cos \b )] \nn \\
& +s_{\th}^2 [2 c_{12} c_{23} (m_1-m_2) s_{12} s_{13} s_{23} \cos \delta + c_{23}^2 ( m_1 s_{12}^2 + m_2 c_{12}^2) + s_{13}^2 s_{23}^2  ( m_1 c_{12}^2 + m_2 s_{12}^2 ) ] \, . 
\end{align}
The maximum value $|m_{11}^{\rm IH}|^{\rm max}$ is exist near the point $\b = \d = \pi$ because of $m_{2} \gtrsim m_{1}$ and terms proportional to $s_{\th}^{2}$ are negligble second-order perturbations.
Thus, the Majorana phase is obviously chosen to be $\a \simeq 0$.

Substituting $m_{1} \simeq m_{2}$, we obtain
\begin{align}
|m_{11}^{\rm IH}|^{\rm max} \simeq m_{1} 
(  c_{\th}^{2} c_{13}^{2} + 2 s_{\th} c_{\th} s_{13} c_{13} s_{23} + s_{\th}^{2} (c_{23}^2+s_{13}^2 s_{23}^2) )  
 \simeq 50 \meV \, . 
\end{align}
The dependence of $\th$ is small because the terms in the first order of $s_{\th}$ cancel out by  $m_{1} \simeq m_{2}$ and the lowest order is proportional to $s_{\th} s_{13}$. 

The minimum value is evaluated in the same manner. 
\begin{align}
& {m_{1} |W_{11}|^{2} - m_{2} |W_{12}|^{2} \over m_{1}}  \\
& \simeq  \abs{c_{\theta } c_{12} c_{13} + s_{\theta } e^{i \b } ( c_{23} s_{12} + c_{12} s_{13} s_{23} e^{i \d})}^{2} 
- \abs{ c_{\theta } s_{12} c_{13} - s_{\theta } e^{i \b } (c_{12} c_{23} - s_{12} s_{13} s_{23} e^{i\d})}^{2} \\ 
& = c_{13}^2 c_{\th}^2  (c_{12}^2-s_{12}^2 ) 
+ 2 c_{13} c_{\theta } s_{\theta } 
[2 c_{12} c_{23} s_{12} \cos \beta + s_{13} s_{23} (c_{12}^2-s_{12}^2 ) \cos (\beta +\delta ) ] \nn \\
& - s_{\theta }^2 [  (c_{12}^2 - s_{12}^2) (c_{23}^2-s_{13}^2 s_{23}^2 ) -4 c_{12} c_{23} s_{12} s_{13} s_{23} \cos \delta ]  \, . 
\end{align}
Among terms with phase dependences, the dominant one is the term proportional to $\cos \b$.
The rests are suppressed by $s_{13}$ and/or $s_{\th}$, and the minimum exists at a point near $\b = \pi, \d = 0$. We have confirmed these behaviors numerically. 
From this, the minimum value $|m_{11}^{\rm IH}|^{\rm min}$ exists around $\a = \pi$,
\begin{align}
|m_{11}^{\rm IH}|^{\rm min} \simeq 
m_{1} (\abs{c_{\theta } c_{12} c_{13} - s_{\theta } ( c_{23} s_{12} + c_{12} s_{13} s_{23} )}^{2} 
- \abs{ c_{\theta } s_{12} c_{13} + s_{\theta } (c_{12} c_{23} - s_{12} s_{13} s_{23} )}^{2}) \, . 
\end{align}
By ignoring the second-order perturbations for $s_{\th}$ and $s_{13}$,
\begin{align}
|m_{11}^{\rm IH}|^{\rm min} & \simeq m_{1} 
(c_{13}^2 c_{\th}^2  (c_{12}^2-s_{12}^2) - 4 s_{\th} c_{\th} s_{12} c_{13}  c_{12} c_{23}) 
 \simeq 19 (1 - 3 s_{\th}) \meV \, . 
\end{align}
The numerical values are approximately
\begin{align}
|m_{11}^{\rm IH}|^{\rm min}_{\th = 0} \simeq 18.6 \meV  \,  , ~~
|m_{11}^{\rm IH}|^{\rm min}_{\th = 0.1} = 12.2 \meV \, ,  ~~
|m_{11}^{\rm IH}|^{\rm min}_{\th = 0.2} = 5.53 \meV \, . 
\end{align}
These expressions coincide with the upper and lower bounds of $|m_{ee}|$ in the limit of $\th \to 0$.

Therefore, in grand unified theories where the chiral perturbative expansion is valid, 
the range of $M_{1}^{\rm IH} = m_{D1}^{2} / |m_{11}^{\rm IH}|$ in IH  with $m_{3} \to 0$ is
\begin{align}
5 \TeV  \lsp {m_{D1} \over 0.5 \, \MeV} \rsp^{2}  
 \lesssim M_{1}^{\rm IH} \lesssim
 50 \TeV \lsp {m_{D1} \over 0.5 \, \MeV} \rsp^{2} \, .
\end{align}
Around these upper and lower bounds, 
all CP phases are almost trivial and some CP symmetry is expected in the lepton sector.
In addition, the latest limit of double beta decay by the KamLAND-Zen collaboration $|m_{ee}| \lesssim 36 - 156 \meV$ \cite{KamLAND-Zen:2022tow} is roughly corresponds to about $7 - 1.6 \TeV \gtrsim M_{1}^{\rm IH}$. 


The parameter $m_{11}$ also dominates $M_{2}$.
From Eq.~(\ref{Mi}) and the behavior~(\ref{minv33}) of $(m^{-1})_{33}$ in SD,
\begin{align}
M_{2}^{\rm IH} \simeq m_{D2}^{2} \abs{m_{11}  \over \Det m (m^{-1})_{33} } 
\simeq {m_{D2}^{2} \over m_{1} m_{2}} \abs{m_{11}  \over  (W_{33}^{*})^{2}   } 
\simeq {m_{D2}^{2} \over m_{1} m_{2}} \abs{m_{11}  \over  U_{\t 3}^{2}   } 
 \, .  
 \label{M2IH}
\end{align}
The fermion masses of the second generation at the GUT scale are roughly 
\cite{Xing:2007fb}
\begin{align}
m_{c} \simeq 200 \MeV , ~~ 
m_{s} \simeq 20 \MeV , ~~
 m_{\m} \simeq 100 \MeV .
\end{align}
By normalizing for $m_{D2} \sim 100 \MeV$, 
\begin{align}
M_{2}^{\rm IH} \simeq 
 \lsp {m_{D2} \over 100 \, \MeV} \rsp^{2} 
 \abs{ m_{11} \over m_{1} } 
 \lsp {10 \, \meV \over m_{2} \, |U_{\t 3}^{2}| } \rsp 
 10^{9} \, \GeV \,  . 
\end{align}
Then the allowed region of $M_{2}^{\rm IH}$ of IH is determined as 
\begin{align}
0.4 \times 10^{9} \GeV  \lsp {m_{D2} \over 100 \, \MeV} \rsp^{2}  \lesssim  M_{2}^{\rm IH} \lesssim 4.0 \times 10^{9} \GeV  \lsp {m_{D2} \over 100 \, \MeV} \rsp^{2}  .
\end{align}
%

\subsection{Normal hierarchy}

We proceed to the analysis of NH. As in the case of IH, there exists the triangular inequality, 
\begin{align}
m_{2} |W_{12}|^{2} - m_{3} |W_{13}|^{2} \leqq  |m_{11}^{\rm NH}| \leqq m_{2} |W_{12}|^{2} + m_{3} |W_{13}|^{2} \, . 
\end{align}
For the upper limit, 
\begin{align}
 & |W_{12}|^{2} m_{2} + |W_{13}|^{2} m_{3} \\
&  \simeq
m_{2} \abs{c_{\theta } s_{12} c_{13} - s_{\theta } e^{i \b } (c_{12} c_{23} - s_{12} s_{13} s_{23} e^{i\d})}^{2} + m_{3} 
\abs{c_{\theta } s_{13} e^{-i \delta } - s_{\theta } e^{i \b} c_{13} s_{23} }^{2} \\
& = 
c_{\th}^2 (m_2 c_{13}^2 s_{12}^2+m_3 s_{13}^2 ) - 
2 s_{\theta } [ c_{13} c_{\th} \{ m_2 c_{12} c_{23} s_{12} \cos \beta + (m_3-m_2 s_{12}^2 ) s_{13} s_{23} \cos (\beta +\delta ) \} ] \nn
\\ & +s_{\theta }^2 [m_2 (c_{12}^2 c_{23}^2+s_{12}^2 s_{13}^2 s_{23}^2) + m_3 s_{23}^2 c_{13}^2 
- 2 m_2 c_{12} c_{23}  s_{12} s_{13} s_{23} \cos \delta ] \, .  \label{mNHlow}
\end{align}
Due to phase-dependent terms of the first-order in $s_{\th}$, 
it reaches the maximum near $\b = \pi , \d = 0$ and $\a = 0$ .
The maximum value $|m_{11}^{\rm NH}|^{\rm max}$ is expressed as 
\begin{align}
|m_{11}^{\rm NH}|^{\rm max} &\simeq 
m_{2} [ c_{\theta } s_{12} c_{13} + s_{\theta } (c_{12} c_{23} - s_{12} s_{13} s_{23} ) ]^{2} + m_{3} (c_{\theta } s_{13} + s_{\theta } c_{13} s_{23} )^{2} \\
 & \simeq 
 ( m_2 s_{12}^2+m_3 s_{13}^2 ) + 2 s_{\theta }  (s_{13} s_{23} ( m_3 - m_2 s_{12}^2  ) +  m_2 c_{12} c_{23} s_{12} ) \, ,
\end{align}
where the second order terms of $s_{\th}$ are ignored in the second equality.
Numerically, it is approximately evaluated as 
\begin{align}
|m_{11}^{\rm NH}|^{\rm max}_{\th = 0} =  3.67 \meV \, ,  ~~
|m_{11}^{\rm NH}|^{\rm max}_{\th = 0.1} =  5.50 \meV\, , ~~
|m_{11}^{\rm NH}|^{\rm max}_{\th = 0.2} = 7.81 \meV \, . 
\end{align}

For $M_{1}^{\rm NH}$, the same equation as IH holds. 
For $M_{2}^{\rm NH}$, we obtain
\begin{align}
M_{2} \simeq m_{D2}^{2} \abs{m_{11}  \over \Det m (m^{-1})_{33} } 
\simeq {m_{D2}^{2} \over m_{2} m_{3}} \abs{m_{11}  \over  U_{\t 1}^{2}   } 
 \, .  
\end{align}
 By a proper normalization, 
\begin{align}
M_{2} \simeq 
 \lsp {m_{D2} \over 100 \MeV} \rsp^{2}
 \lsp { 100 \meV^{2} \over m_{2} m_{3} |U_{\t 1}|^{2}}  \rsp
 \lsp { |m_{11}|  \over 1 \meV } \rsp
  10^{8} \, \GeV \,  . 
\end{align}
Therefore, the minimum value of $M_{1}^{\rm NH}$ and the maximum value of $M_{2}^{\rm NH}$ are given by the maxmum value of $|m_{11}|$ with $\th = 0.2$, 
\begin{align}
M_{1}^{\rm min} \simeq 30 \TeV  \lsp {m_{D1} \over 0.5 \, \MeV} \rsp^{2} \, ,
~~ 
M_{2}^{\rm max} \simeq 3 \times 10^{8} \GeV \lsp {m_{D2} \over 100 \, \MeV} \rsp^{2} \, . 
\end{align}

On the other hand, the minimum of $m_{11}^{\rm NH}$ is
\begin{align}
 & |W_{12}|^{2} m_{2} - |W_{13}|^{2} m_{3} \\
& \simeq
m_{2} \abs{c_{\theta } s_{12} c_{13} - s_{\theta } e^{i \b } (c_{12} c_{23} - s_{12} s_{13} s_{23} e^{i\d})}^{2} - m_{3} 
\abs{c_{\theta } s_{13} e^{-i \delta } - s_{\theta } e^{i \b} c_{13} s_{23} }^{2} \\
& = m_2 s_{12}^2 c_{13}^2 c_{\th}^2 - m_3(  c_{13}^2 s_{23}^2 s_{\th}^2 + s_{13}^2 c_{\th}^2 )
+ 2 c_{13} c_{\theta } s_{\theta } \left(s_{13} s_{23} \left(m_2 s_{12}^2+m_3\right) \cos  (\b + \d ) - c_{12} c_{23} m_2 s_{12}  \cos \b \right) \nn \\
 & +m_2 s_{\theta }^2 (c_{12}^2 c_{23}^2+s_{12}^2 s_{13}^2 s_{23}^2 -2 c_{12} c_{23} s_{12} s_{13} s_{23} \cos \d ) \,  .
\end{align}
This is minimized near $\b = 0$ and $\d = \pi$.
Thus the minimum value $(m_{11}^{\rm NH})^{\rm min}$ of $m_{11}^{\rm NH}$ exists near $\a \simeq \pi$, and a rough evaluation of the first order of $s_{\th}$ is
\begin{align}
(m_{11}^{\rm NH})^{\rm min}
&\simeq  m_{2} (c_{\theta } s_{12} c_{13} - s_{\theta }  (c_{12} c_{23} + s_{12} s_{13} s_{23} )) ^{2} - m_{3} 
(c_{\theta } s_{13} + s_{\theta }  c_{13} s_{23})^{2} \\
& \simeq 
 (m_2 s_{12}^2 - m_3 s_{13}^2) - 2  s_{\theta } (s_{13} s_{23} (m_3 + m_2 s_{12}^2 ) +m_2  c_{12} c_{23} s_{12} ) \, . 
\end{align}
The numerical values are  
\begin{align}
(m_{11}^{\rm NH})^{\rm min}_{\th = 0} \simeq 1.44 \meV  \,  , ~~
(m_{11}^{\rm NH})^{\rm min}_{\th = 0.1} = -0.50 \meV \,  , ~~
(m_{11}^{\rm NH})^{\rm min}_{\th = 0.2} = -2.92 \meV \, . 
\end{align}

For a finite $s_{\th}$, $m_{11}^{\rm NH}$ can have a negative value. 
This corresponds to the region where $|m_{11}^{\rm NH}|^{\rm min} = 0$ is possible for appropriate complex phases.
The condition for such a cancellation to occur is $\sqrt{m_{2}} W_{12} \simeq \pm i \sqrt{m_{3}} W_{13}$, and the minimum $\th$ that satisfies the condition is
\begin{align}
 \sqrt {m_{2}} (c_{\theta } s_{12} c_{13} - s_{\theta }  (c_{12} c_{23} + s_{12} s_{13} s_{23} )) = 
 \pm  \sqrt{m_{3}} (c_{\theta } s_{13} + s_{\theta }  c_{13} s_{23} ) \, . 
\end{align}
By choosing the solution with the smaller $\th$, 
\begin{align}
s_{\theta } = 
 \frac{c_{13} s_{12}-\sqrt{\frac{m_3}{m_2}} s_{13}}{c_{13} \sqrt{\frac{m_3}{m_2}} s_{23}+c_{12} c_{23}+s_{12} s_{13} s_{23}}
 \simeq 0.077 \, . 
\end{align}
For a larger $\th$ than this value, there is a canceling region where $(m_{11}^{\rm NH})^{\rm min}\simeq 0$.
The phase dependence of the minimum is shown in Fig.~1. 
We can see that such cancellation is likely to occur in the region $\d + \b \sim \pi $.

\begin{figure}[h]
\begin{center}
\begin{tabular}{cc}
 \includegraphics[width=6cm]{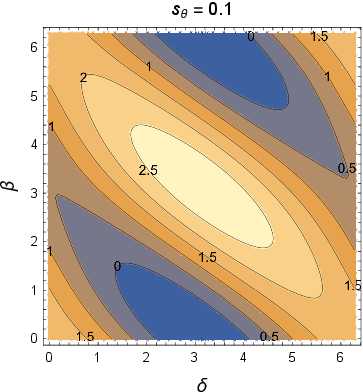} ~&~ 
 \includegraphics[width=6cm]{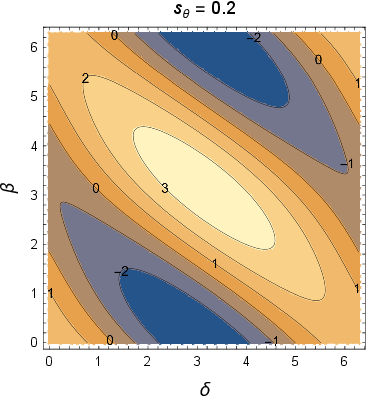} 
\end{tabular}
\caption{ 
Plots of $ |W_{12}|^{2} m_{2} - |W_{13}|^{2} m_{3}$~(\ref{mNHlow}) for $\d$ and $\b$ in NH. The left and right ones represent  the case of $s_{\th} = 0.1, 0.2$ respectively. }
\end{center}
\end{figure}
%

\subsection{$M_{1,2}$ in the level crossing point $(m_{11}^{\rm NH})^{\rm min} \simeq 0$}

The chiral perturbative expansion breaks down in the limit of $m_{11} \to 0$. This is clear from Eq.~(\ref{MR0}) because the diagonalization of the mass matrix $M_{R0}$ for the lighter generations has a large mixing.
In this {\it level crossing point} \cite{Akhmedov:2003dg}, the masses $M_{1}$ and $M_{2}$ are of equal magnitude, and the relation $M_{1}^{-1} \propto m_{11} \propto M_{2}$ no longer holds.
Let us examine how the mass matrices $m$ and $M_{R0}$ behave in the region where $m_{11}$ is small.
From Eq.~(\ref{minv33}), the minor determinant is evaluated as 
\begin{align}
| m_{11} m_{22} - m_{12}^{2} | = | \Det m  (m^{-1})_{33} |
\simeq  m_{2} m_{3} |{(W_{31}^{*})^{2} }|
\simeq  m_{2} m_{3}  |U_{\t 1}|^{2}  \, . 
\end{align}
Thus in the limit of $m_{11} \to 0$, $|m_{12}| \simeq  \sqrt {m_{2} m_{3}}  |U_{\t 1}| \simeq 15 \meV$ holds. 
Since the value $|m_{22}| \simeq |m_{\m \m}| \simeq m_{3} /2 \simeq 25 \meV$ is a roughly constant under $\th$,   
the behaviors of $m$ and $M_{R0}$~(\ref{MR0}) in the limit of $m_{11} \to 0$ are
\begin{align}
M_{R0} & =  {1\over m_{11} m_{22} - m_{12}^{2}}
\diag{m_{D1}}{m_{D2}}
\begin{pmatrix}
m_{22} & - m_{12} \\ - m_{12} & m_{11}
\end{pmatrix} 
\diag{m_{D1}}{m_{D2}} \, , \\
|M_{R0}| & \simeq  {1\over  m_{2} m_{3}  |U_{\t 1}|^{2}}
\begin{pmatrix}
m_{D1}^{2} m_{3} /2 & m_{D1} m_{D2} \sqrt {m_{2} m_{3}}  |U_{\t 1}| \\ 
m_{D1} m_{D2}  \sqrt {m_{2} m_{3}}  |U_{\t 1}| & 0
\end{pmatrix}  .
\end{align}
From $|m_{12}| \sim |m_{22}|$, 
the off-diagonal element is dominant in unified theories with $m_{D2} \gtrsim 20 \, m_{D1}$, 
and  the two mass singular values are almost same $M_{1} \simeq M_{2}$.

Such a condition is also evaluated from the relation of determinants and Eq.~(\ref{13}); 
\begin{align}
M_{1} M_{2} M_{3} = {m_{D1}^{2} m_{D2}^{2} m_{D3}^{2} \over m_{1} m_{2} m_{3}} \, , ~~~ 
M_{1} M_{2}  = {m_{D1}^{2} m_{D2}^{2}  \over  m_{2} m_{3} |U_{\t 1}|^{2}}  \, .
\end{align}
When $M_{1}$ and $M_{2}$ are approximately equal, 
\begin{align}
M_{1} \simeq M_{2} & \simeq {m_{D1} m_{D2} \over \sqrt{m_{2} m_{3}} |U_{\t 1}| } \, . 
\end{align}
It gives an upper bound on $M_{1}^{\rm NH}$ and a lower bound on $M_{2}^{\rm NH}$ simultaneously,
\begin{align}
M_{1}^{\rm max} \simeq M_{2}^{\rm min} \simeq 
 \lsp {m_{D1} \over 1 \MeV} \rsp 
 \lsp {m_{D2} \over 100 \MeV} \rsp 
 \lsp { 10 \meV \over |U_{\t 1}| \sqrt{ m_{2} m_{3}} } \rsp
  10^{7} \, \GeV \,  . 
\end{align}
Therefore, the bounds will be about $10 \PeV$. 
Unlike with the case of IH, these upper and lower bounds are proportional to $m_{D1} m_{D2}$.
To summarize these results, 
\begin{align}
3 \times 10^{4} \GeV \lsp {m_{D1} \over 0.5 \, \MeV} \rsp^{2} &\lesssim M_{1}^{\rm NH} \lesssim 3 \times 10^{6} \GeV \lsp {m_{D1} m_{D2} \over 50 \, \MeV^{2}} \rsp  ,  \\
3 \times 10^{6} \GeV \lsp {m_{D1} m_{D2} \over 50 \, \MeV^{2}} \rsp &\lesssim M_{2}^{\rm NH} \lesssim 3 \times 10^{8} \GeV \lsp {m_{D2} \over 100 \, \MeV} \rsp^{2}  . 
\end{align}
Since $M_{1} M_{2}$ is approximately constant under $\th$, the widths of these allowed regions are comparable.

\section{Summary}

In this paper, we perform a chiral perturbative evaluation of the (second) lightest mass $M_{1,2}$ of right-handed neutrino $\n_{R1,2}$ in the type-I seesaw mechanism with $SO(10)$-inspired relations and an almost massless neutrino $m_{1 \, \rm or \, 3} \sim 0$. 
By chiral perturbative treatment, 
the masses $M_{1,2}$ are expressed as $M_{1} = m_{D1}^{2}/m_{11} , \, M_{2} = m_{D2}^{2} m_{11} / (m_{11} m_{22} - m_{12}^{2})$ with the mass matrix of left-handed neutrinos $m$ in the diagonal basis of the Dirac mass matrix $m_{D}$.
Assuming $m_{Di}$ and the unitary matrix $V$ in the singular value decomposition $(m_{D})_{ij} = \sum_{k} V_{ik} m_{D k} U^{\dg}_{kj}$ are close to observed fermion masses and the CKM matrix, we obtained expressions for $M_{1,2}$ by parameters in the low energy and unknown phases. 

As a result, the upper and lower bounds of $M_{1,2}$ are obtained as
\begin{align}
3 \times 10^{4} \GeV \lsp {m_{D1} \over 0.5 \, \MeV} \rsp^{2}  &\lesssim M_{1}^{\rm NH} \lesssim 3 \times 10^{6} \GeV  \lsp {m_{D1} m_{D2} \over 50 \, \MeV^{2}} \rsp ,  \\
3 \times 10^{6} \GeV  \lsp {m_{D1} m_{D2} \over 50 \, \MeV^{2}} \rsp &\lesssim M_{2}^{\rm NH} \lesssim 3 \times 10^{8} \GeV \lsp {m_{D2} \over 100 \, \MeV} \rsp^{2}   .
\end{align}
for the NH, and 
\begin{align}
5 \times 10^{3} \GeV  \lsp {m_{D1} \over 0.5 \, \MeV} \rsp^{2} &\lesssim  M_{1}^{\rm IH} \lesssim 5 \times 10^{4} \GeV  \lsp {m_{D1} \over 0.5 \, \MeV} \rsp^{2}  \, ,  \\
4 \times 10^{8} \GeV   \lsp {m_{D2} \over 100 \, \MeV} \rsp^{2}  &\lesssim  M_{2}^{\rm IH} \lesssim 4 \times 10^{9} \GeV   \lsp {m_{D2} \over 100 \, \MeV} \rsp^{2}  \, ,
\end{align}
for the IH. 
The difference in the dependence of $m_{Di}$ is due to a cancellation of $m_{11}$ in NH.
This result would be valid for discussions of cosmology, such as leptogenesis, because it is encompassed by the results of some grand unified theories with seesaw mechanisms. 


\end{document}